\documentstyle[12pt]{article}

\def\be{\begin{equation}}
\def\ee{\end{equation}}
\def\ba{\begin{eqnarray}}
\def\ea{\end{eqnarray}}

\def\R{{\cal R}}

\def\agb{{\overline {{\cal A}/{\cal G}}}}

\def\C{{\cal C}}

\def\Comp{{\mathchoice
{\setbox0=\hbox{$\displaystyle\rm C$}\hbox{\hbox to0pt
{\kern0.4\wd0\vrule height0.9\ht0\hss}\box0}}
{\setbox0=\hbox{$\textstyle\rm C$}\hbox{\hbox to0pt
{\kern0.4\wd0\vrule height0.9\ht0\hss}\box0}}
{\setbox0=\hbox{$\scriptstyle\rm C$}\hbox{\hbox to0pt
{\kern0.4\wd0\vrule height0.9\ht0\hss}\box0}}
{\setbox0=\hbox{$\scriptscriptstyle\rm C$}\hbox{\hbox to0pt
{\kern0.4\wd0\vrule height0.9\ht0\hss}\box0}}}}
\def\Co{{\mathchoice
{\setbox0=\hbox{$\displaystyle\rm C$}\hbox{\hbox to0pt
{\kern0.4\wd0\vrule height0.9\ht0\hss}\box0}}
{\setbox0=\hbox{$\textstyle\rm C$}\hbox{\hbox to0pt
{\kern0.4\wd0\vrule height0.9\ht0\hss}\box0}}
{\setbox0=\hbox{$\scriptstyle\rm C$}\hbox{\hbox to0pt
{\kern0.4\wd0\vrule height0.9\ht0\hss}\box0}}
{\setbox0=\hbox{$\scriptscriptstyle\rm C$}\hbox{\hbox to0pt
{\kern0.4\wd0\vrule height0.9\ht0\hss}\box0}}}}
\def\Rl{{\mathchoice
{\setbox0=\hbox{$\displaystyle\rm R$}\hbox{\hbox to0pt
{\kern0.4\wd0\vrule height0.9\ht0\hss}\box0}}
{\setbox0=\hbox{$\textstyle\rm R$}\hbox{\hbox to0pt
{\kern0.4\wd0\vrule height0.9\ht0\hss}\box0}}
{\setbox0=\hbox{$\scriptstyle\rm R$}\hbox{\hbox to0pt
{\kern0.4\wd0\vrule height0.9\ht0\hss}\box0}}
{\setbox0=\hbox{$\scriptscriptstyle\rm R$}\hbox{\hbox to0pt
{\kern0.4\wd0\vrule height0.9\ht0\hss}\box0}}}}

\def\C{{\cal C}}

\def\o{\overline}

\def\ab{\o {\cal A}}

\def\C*{$C^{\star}$}
\def\ge{\geq}

\let\ssection=\section
\renewcommand{\section}{\setcounter{equation}{0}\ssection}

\title{The Status of Diffeomorphism Superselection in Euclidean 2+1 Gravity}
\author{Donald Marolf\thanks{Department of Physics, Syracuse University,
Syracuse, NY 13244, USA},
Jos\'e Mour\~ao\thanks{Departamento de F\'{\i}sica, Instituto Superior
Tecnico, Av. Rovisco Pais, 1096 Lisboa Codex, PORTUGAL},
Thomas Thiemann\thanks{Physics Department, Harvard University, Cambridge,
MA 02138, USA, Internet:thiemann@math.harvard.edu}}
\date{\today}

\begin{document}


\maketitle


\begin{abstract}
This work addresses a specific technical question of relevance to
canonical quantization of gravity using the so-called new variables 
and loop-based techniques of Ashtekar, Rovelli, and Smolin.  In 
particular, certain `superselection laws' that
arise in current applications of these techniques to solving the
diffeomorphism constraint are considered.  Their status is elucidated by
studying an analogous system: 2+1 Euclidean gravity.  For that
system, these superselection laws are shown to be spurious.  This, however,
is only a technical difficulty.  The usual quantum theory may still
be obtained from a loop representation and the technique known as
`Refined Algebraic Quantization.'

\end{abstract}
\begin{section} {Introduction}
\label{intro}

A recent advance in canonical quantization techniques was the introduction
\cite{jmp}
of Refined Algebraic Quantization (and related techniques
\cite{Landsman,Hig}) for solving quantum
constraints and for inducing physical inner products.  As shown in
\cite{jmp} the use of such techniques often results in `superselection
rules.'  While such superselection rules can correspond 
to important properties of the physical system \cite{jmp,ban}
which are present even
at the classical level, when RAQ is used to solve the diffeomorphism
constraints of a quantum theory of connections as in \cite{jmp}, 
the interpretation of the superselection rules is less clear.

In particular, when 2+1 gravity is expressed as a theory of
connections \cite{Witten,Mon,Mess}, the simplest observables appear to
violate these rules.  This is because, in a loop representation, 
these rules select between states associated with different topological
types of graphs or loops, while important observables in the 2+1
theory are traces of holonomies of connections around noncontractable
curves, which mix the above states. 
 
In a loop representation, such operators
change not only the topology but
also the {\it homotopy} type of a loop-state.  The goal of this
paper is to determine the status of these superselection rules
in Euclidean 2+1 gravity and to determine whether their presence
prevents the quantization scheme described in \cite{jmp} from
succeeding.  This will help to clarify the standing of such methods in
the loop-based approach to 3+1 gravity.

We will proceed in two stages.  It will first be shown
that methods based on a loop
representation and
the `Refined Algebraic Quantization scheme' (RAQ)
of \cite{jmp} {\it do}
yield the usual results \cite{Witten,AAbook,JR} for Euclidean
2+1 gravity when they are properly applied.  In this case, 
the most straightforward treatment 
differs from the particular approach suggested in 
\cite{jmp}.   However, we also show that the
solution may be recast in the form advocated in \cite{jmp}
in which the diffeomorphism constraints
are solved first and the Hamiltonian constraints are then solved
as a second stage.  Regarding the `diffeomorphism superselection
rules' mentioned above, we will see that they disappear
in the final solution of this system.  Some concluding comments about
expectations for the 3+1 theory are given in section \ref{concl} and we draw on
supporting material from an appendix.  This discussion
suggests that the intermediate presence of the superselection rules in the 
2+1 theory is due to the singular nature of our description of the 
theory, but that a similar singularity may be present in the loop
approach to the  3+1 case \cite{jmp}.

This work will make use of a loop representation along the lines of
\cite{jmp} as well as the Refined Algebraic techniques discussed
there, in \cite{ban}, and elsewhere.  As a result, what follows is
best considered a technical addendum to \cite{jmp} and the review of
that material will be kept to a minimum.   We use the same
structures and definitions as \cite{jmp}, except as where noted below.
We will, however, briefly discuss the formulation of Euclidean
2+1 gravity as a canonical theory of connections since that 
was not discussed in \cite{jmp}.

As described by Witten
\cite{Witten}, Euclidean 2+1 vacuum gravity may be considered
as a theory of co-triads $\overline{e}_{aI}$ and $SU(2)$ valued
connections $\overline{A}_a^I$.  Here, $a,b$ are spacetime indices on a three
manifold $M$.  The system is governed by the action
\begin{equation}
S(\overline{e}_{aI}, \overline{A}_a^I) =  {1 \over 2}
\int_M d^3x \tilde{\epsilon}^{abc} \overline{e}_{aI} \overline{F}^I_{bc}
\end{equation}
where $\tilde{\epsilon}^{abc}$ is the Levi-Civita density on $M$ and 
$\overline{F}^I_{bc}$ is the curvature of the connection $\overline{A}^I_a$.
This is just the 2+1 Einstein-Hilbert action written in terms of the
triad and spin connection.  For later convenience we have taken the action
to differ from that of \cite{Witten} by a factor of $1 \over 2$. 

If we now take $M$
to be of the form ${\bf R} \times \Sigma$ (for a closed orientable
two-manifold $\Sigma$), we may make a 2+1 decomposition of the above
action.  The result is a system where the
Hamiltonian is simply a sum of constraints.  We shall take
$i,j,k$ to be abstract indices associated with the manifold
$\Sigma$.  The canonical variables are a connection $A^I_i$ which
is the pull back of the connection $\overline{A}^I_a$ to $\Sigma$
and a vector density $\tilde{E}_I^i =
\tilde{\epsilon}^{ij} e_{jI}$ where $\epsilon^{ij}$ is the Levi-Civita
density on $\Sigma$ and $e_{jI}$ is the pull back of $e_{aI}$ to 
$\Sigma$.  These satisfy the canonical commutation relations
\begin{equation}
\Big\{ A_i^I(x), \tilde{E}^j_I(x') \Big\} = \delta^j_i \delta^I_J \delta^2(x,x')
\end{equation}
and, in terms of $A^I_i, \tilde{E}^j_I$, the constraints are
\begin{equation}
\label{WC}
F^I_{ij} = 0 \ \ D_j \tilde{E}^j_I = 0
\end{equation}
where $F^I_{ij}$ and $D_j$ are the curvature and covariant derivative 
associated with $A_i^I$ respectively.  

The second constraint is known
as the Gauss constraint and generates $SU(2)$ gauge transformations.
The first constraint is more complicated, but clearly generates
transformations that do not change the connection.  The reader will, at
this point, notice the distinct lack of a constraint that generates
diffeomorphisms.  Such a constraint would have the form $\tilde{E}^i_I
F^I_{ij} = 0$.  Although it is not one of the 
constraints (\ref{WC}), this function clearly vanishes on the constraint
surface; the result is that any function invariant under the transformations
generated by the $F = 0$ constraint also becomes invariant under
diffeomorphisms once it has been restricted to the constraint
surface.  In this sense then, the
Witten constraints are in fact weakly equivalent (for non-degenerate
triads) to the set of constraints \cite{Beng} 
\begin{equation}
\label{BC}
D_i \tilde{E}^i_I = 0, \ \ \tilde{E}^i_I F^I_{ij} = 0, \ \  
\epsilon^{IJ}_K \tilde{E}^i_I \tilde{E}^j_J F^K_{ij} = 0,
\end{equation}
but (\ref{WC}) and (\ref{BC}) are not strongly equivalent.  

We are therefore left with the question of which set of constraints
to use here.  On the one hand, the well understood description of
2+1 gravity refers to the constraints (\ref{WC}).  On the other, 
we are most interested in gaining insight into 3+1 gravity, for which
only a description of the form (\ref{BC}) is available.  Furthermore, 
the question of the `diffeomorphism superselection sectors' which
we wish to study does not arise unless there is in fact a diffeomorphism
constraint. However, the densitized `Hamiltonian 
constraint' $EEF = 0$ of (\ref{BC}) is as difficult to define here
as in the 3+1 case.

One approach might be to apply techniques such as those
introduced by Thiemann for the 
3+1 Hamiltonian constraints in \cite{TT}.   However,
because of conceptual and technical 
complications involved, we leave direct investigation
of the constraints (\ref{BC}) for future work
\cite{AL1} and content 
ourselves here with following a hybrid approach.  After briefly
reviewing the refined algebraic techniques in section 2, we define
our system using the Witten constraints (\ref{WC}) and show that the 
combination of a loop representation with   Refined Algebraic techniques 
generates the usual physical Hilbert space in a straightforward manner.
In section 3, we show that the physical states generated in this way
{\it are} annihilated by the diffeomorphism constraint 
$\tilde{E}^i_I F^I_{ij} = 0$ in the sense described in \cite{jmp} and
that our physical Hilbert space could have
been constructed by following the procedure outlined in \cite{jmp}, 
in which the diffeomorphism constraint $\tilde{E}^i_I F^I_{ij}$ is
solved first (through RAQ) and the `remaining parts' of the
constraints (\ref{WC}) are solved later by RAQ-like techniques.
Section 4 discusses the implications for the 3+1 theory, 
drawing on the appendix for support.

\end{section}
\begin{section} {Quantization}

We now proceed to quantize the system described in section \ref{intro}
and to impose the constraints (\ref{WC}) using a loop representation
and the techniques of Refined Algebraic quantization.  That is to 
say, we will follow \cite{jmp} in considering an auxiliary kinematical
Hilbert space $H_{aux} = L^2(\agb,d\mu_0)$ where $\agb$ is the 
Ashtekar-Isham `quantum configuration space of gauge equivalent
connections' \cite{AI} 
appropriate to the connections
discussed above.  This space contains not only connections but
suitably generalized `distributional' connections as might be
expected to be required in the configuration space of a quantum field
theory.  Note that $d\mu_0$ is the corresponding Ashtekar-Lewandowski
measure \cite{AL}.  States in this space are gauge invariant, 
so there is no need to impose the Gauss constraints\footnote{If one
wishes, one may \cite{jmp} begin with a larger Hilbert space
$L_2(\ab,d\hat{\mu}_0)$ whose states are not gauge invariant, introduce
the Gauss constraints as operators on this space, and solve them
by RAQ to arrive at $H_{aux} = L^2(\agb,d\mu_0)$ as above.}; they
are considered to be identically satisfied on this space.

We must, however, define and solve the $F=0$ constraints.  
This involves a slight complication as the generalized 
connections of $\agb$ are not in general differentiable.  Thus, 
the curvature $F$ is strictly speaking not well defined on this space.
What {\it is} well-defined though is the holonomy of a generalized 
connection -- this is in fact the very definition of $\agb$.
We therefore proceed
as follows : the statement $F=0$ for a smooth connection $A$ is
equivalent to the statement that the holonomy $h_\alpha(A)$ of $A$ around 
each contractable loop $\alpha$
in $\Sigma$ is trivial, that is, gives just the unit element of $SU(2)$.
A manifestly gauge invariant formulation of the constraints is thus
\be \label{FC1}
C'_\alpha:=2-T_\alpha(A)=0
\ee
where $T_\alpha(A):=\mbox{tr}(h_\alpha(A))$, for all contractable loops.
The virtue of writing the $F=0$ constraint in the form (\ref{FC1}) is that
we can extend it to $\agb$. The disadvantage is that this constraint 
classically does not generate gauge transformations on the constraint 
surface as $T_\alpha-2$ is quadratic in $F$.   We will
actually use a constraint of the form
\begin{equation}
\label{FC2}
C_{\alpha} = |2-T_\alpha(A)|^{3/2}
\end{equation}
for all contractable loops $\alpha$.  While, classically, this is
even worse than (\ref{FC1}), we shall see that a certain
`cancellation of singularities' occurs, and that this is in fact
a preferred form of the constraints.  

We now wish to solve these constraints using the Refined Algebraic
quantization procedure.  Recall that this involves introducing
a preferred dense subspace $\Phi$ of the Hilbert space $H_{aux}$ and
defining an (anti-linear)  map $\eta$ from $\Phi$ to its dual
$\Phi'$ which `projects a state onto the constraint surface' in the
sense that the image $\eta \phi$ of any $\phi \in \Phi$ is a solution
of the constraints: $ C (\eta \phi) = 0$, where $C$ denotes the 
constraints.  Note that the action of $C$ on $\Phi'$ is defined to be
the dual of its action on $\Phi$.  The details are given in
\cite{jmp}, but we remind the reader that the solutions $\eta \phi$
in the image of $\eta$ are given a Hilbert space structure
through
\begin{equation}
\label{ip}
\langle \eta \alpha | \eta \beta \rangle \equiv 
(\eta \beta) [\alpha].
\end{equation}
The map $\eta$ must be real and positive in the sense that, for all
$\phi_1,\phi_2 \in \Phi$, 
\begin{equation}
(\eta \phi_1)[\phi_2] = ((\eta \phi_2)[\phi_1])^*
\ \ \rm{and} \ \ (\eta \phi_1)[\phi_1] \ge 0
\end{equation}
and $\eta$ must commute with every strong observable $A$.  That is, 
for any operator $A$ which commutes with all gauge transformations, 
we must have
\begin{equation}
(\eta \phi_1) [A \phi_2] = ((\eta A^\dagger \phi_1))[\phi_2].
\end{equation}
In this case (\ref{ip}) defines an inner product which may be used to
complete the set of states $\eta \phi$ in the image of $\eta$ to
a `gauge invariant' Hilbert space $H_{Inv}$.  Moreover,
this inner product has the property that any strong observable $A$
on $H_{aux}$ induces on operator $A_{Inv}$ on the physical Hilbert 
space satisfying the same reality conditions; i.e. $A_{Inv}^\dagger
= (A^\dagger)_{Inv}$.  The operator $A_{Inv}$ is defined by
\begin{equation}
A_{Inv} (\eta \phi ) = \eta(A \phi).
\end{equation}
Note that the invariant Hilbert space was referred to as the 
`physical' Hilbert space in \cite{jmp,ban}.  The terminology we
use here is more appropriate for the current setting, in which we allow
the possibility that this procedure be applied more than once, solving
only some of the constraints at each step.

A nice idea for constructing the map $\eta$ is through `group
averaging' \cite{jmp,Hig,ban}.  Under appropriate conditions, an
expression of the form
\begin{equation}
\label{groupav}
(\eta \phi_1)[\phi_2] = \int_G dg \langle \phi_1 | U(g) | \phi_2 \rangle,
\end{equation}
with $dg$ the Haar measure on the gauge group $G$,
gives a well-defined map $\eta$ with the required properties.
This heuristic idea is often quite useful in applying RAQ, although
it will not be of direct use for our case.

The constraint $F=0$ is a pure configuration constraint : it does not 
involve the canonical
momenta. This situation is reminiscent of solving the 
relativistic free particle constraint
$p^2+m^2=0$ in the momentum representation. Let us
recall how this works as it will clarify our case.

In the relativistic particle case we choose $H_{aux}:=L_2(\Rl^2,d^4p)$ 
and $\Phi:=C^\infty_0(\Rl^4)$, say, the smooth test functions of 
compact support.  The constraint 
$C=p^2+m^2=0$ is easy to solve: each solution can be written
in the form $\psi_f(p)=\delta(C)
f(\vec{p})$ where $f\in \Phi$. The point is that $\psi_f\in\Phi'$
is not an element of $H_{aux}$. But why can we claim that the 
constraint was solved by group averaging ? This is because 
$\hat{C}:=\hat{p}^2+m^2$ is an essentially self-adjoint operator on $H_{aux}$
with core $\Phi$ whose unique self-adjoint extension we may exponentiate to 
obtain a 1-parameter 
unitary group $\hat{U}(a):=\exp(ia\hat{C})$ with $a\in\Rl$. The Haar measure
on $\Rl$ is the Lebesgue measure and so for each $\phi\in\Phi$ we obtain 
the following group average map
\be\label{ga}
(\eta f)(\phi):=\int_\Rl \frac{da}{2\pi} <f,\hat{U}(a)\phi>=\psi_f(\phi)
\ee
in analogy with (\ref{groupav}). 

Why does this lead to the desired result ? 
The answer is that one way of looking at the group average 
procedure is that one wishes to solve the exponentiated constraint 
$\hat{U}(a)=1\forall a$ and the average over all the $\hat{U}(a)$ has to 
be done in such a way that we get the $\delta(\hat{C})$ back, or, in 
other words, such that the translation of the parameter $a$ in 
$\hat{U}(a)\hat{U}(b)=\hat{U}(a+b)$ is irrelevant because we are using 
a translation invariant measure (the Haar measure) on the parameter 
space. 

Another feature of the relativistic particle shared by our model is  
that the solutions to $C=0$ are not unique. For the free particle they
are two-fold,
$p^0=\pm\sqrt{\vec{p}^2+m^2}=:\pm\omega(\vec{p}$ which we may encode in the 
following way $C=C_+ C_-$ where $C_\pm=p^0\pm\sqrt{\vec{p}^2+m^2}$. Also, 
in the sense of distributions $\delta(C)=\frac{1}{2\omega}[\delta(C_+)
+\delta(C_-)]=:\int_{\cal M} d\nu(\omega) \delta(p^0,\omega)$ where 
${\cal M}=\{\pm\omega(\vec{p})\}$ is the solution space and $\nu$ is 
proportional to a counting measure on $\cal M$. We will encounter 
precisely the same structure in our model.
This concludes the discussion of the relativistic particle.

Let us now turn to our case. The solutions to $F=0$ are the flat connections,
and, since we are interested only in gauge-invariant information, we
have the space $\cal M$, the moduli space of flat connections modulo 
gauge transformations as our solution space. Therefore, we write the
distribution $\delta(F)$ as 
\be \label{modspace}
\delta(F)=\int_{\cal M} d\nu(A_0) \delta(A,A_0)
\ee
where $\nu$ is some (real-valued)
measure on $\cal M$.  We will derive a preferred measure $d\nu$ below which
agrees with the one give by Witten 
\cite{Witten}.
This 
is in direct analogy with
writing the $\delta(p^2+m^2)$ as a sum of two $\delta$ distributions, the 
discrete measure there was replaced by the measure $\nu$ accounting for 
the fact that $\cal M$ is a manifold.

The next step is non-trivial : we have to write $\delta(A_0,A)$ as a 
well-defined distribution on a suitable $\Phi$. Let us choose, as in the 
$3+1$ case, $H_{aux}:=L_2(\agb,d\mu_0)$ where $\mu_0$ is the 
Ashtekar-Lewandowski measure and let $\Phi:=\Phi_{Cyl}$ be the cylindrical 
functions on $\agb$. It turns out that $H_{aux}$ has an orthonormal basis,
the so-called spin-network states $T_{\gamma,\vec{j},\vec{c}}$ (see
\cite{jmp}). Here 
$\gamma$ stands for a piecewise analytic closed graph, 
$\vec{j}=(j_1,..,j_E)$ is a labeling of its edges $e_1,..,e_E$ with spin
quantum numbers and $\vec{c}=(c_1,..c_V)$ is a labeling of its vertices 
with certain $SU(2)$ invariant matrices. The state 
$T_{\gamma,\vec{j},\vec{c}}$ is built from $\vec{c}$ and 
$\otimes_{k=1}^E \pi_{j_k}(h_{e_k}(A))$, where $\pi_j$ is the $j-th$ 
irreducible representation of $SU(2)$, by contraction of all group indices in 
such a way that it is gauge invariant.
We may use such states to represent $\delta(A,A_0)$ as:
\be \label{distr} 
\delta(A_0,A)=\sum_{\gamma,\vec{j},\vec{c}} 
\overline{T_{\gamma,\vec{j},\vec{c}}(A)}T_{\gamma,\vec{j},\vec{c}}(A_0)
\ee
since, by the orthonormality of spin networks, 
this satisfies $\int d\mu_0(A) \phi(A) = \phi(A_0)$ for all
$\phi \in \Phi_{Cyl}$.  The associated rigging map $\eta_F: \  \Phi_{Cyl}
\rightarrow \Phi'_{Cyl}$ 
is given by
\begin{equation} \label{rigger}
(\eta_F \psi) (\phi ) = \int_{\agb} d\mu_0(A) \overline{\psi(A)}
\overline{\delta(F)} \phi(A).
\end{equation}
Notice that, although the sum $(\ref{distr})$ ranges over a complete set of
piecewise analytic 
graphs (an uncountable set), the result $\eta_F \psi$ is still 
a well-defined element of $\Phi_{cyl}'$.

Can the result (\ref{rigger}) also be obtained by explicitly averaging 
the constraints (\ref{FC1}) in analogy with (\ref{ga}) ? At least
at a heuristic level, the answer is in 
the affirmative\footnote{At a more technical level, there is a subtlety in
that the group generated by the full set of constraints (\ref{FC2})  
does not fit well with the projective structure of ${\cal H}_{aux}$.}.
To see this, notice first that one can write the 
delta distribution on $SU(2)$ with respect to the Haar measure $\mu_H$
as follows 
\be \label{dd}
\delta(g,1)=\int_\Rl \frac{dt}{2\pi}\exp(it[1-\mbox{tr}(g)/2]^{3/2})
\ee
as the reader can check himself by explicitly writing $\mu_H$ in terms of 
local coordinates on $S^3$.  Note that the power $3/2$ is important here
as it cancels certain singularities (actually, degeneracies) in the measure.
This observation motivates us to construct a 
cylindrical definition of $\eta_F$ which we sketch below :\\
For each graph $\gamma$ choose a set of generators $\alpha_1(\gamma),..,
\alpha_{n(\gamma)}(\gamma)$ of the subgroup of the homotopy group of 
$\gamma$ corresponding to contractable loops on $\Sigma$.
Let now
\be \label{ga1}
U_\gamma(t_1,..,t_{n(\gamma)}):=\prod_{i=1}^{n(\gamma)}
U_{\alpha_i(\gamma)}(t_i)\mbox{ where }
U_\alpha(t):=\exp(itC_\alpha).
\ee
We are now in the position to define $\eta_F$ cylindrically : since each 
$\phi,\psi\in\Phi_{cyl}$ are just finite linear combinations of 
spin-network states it will be sufficient to define $\eta_F$ on 
spin-network states $\psi=T_{\gamma,\vec{j},\vec{c}}$ through (\ref{rigger})
for each $\phi=T_{\gamma',\vec{j}',\vec{c}'}$. It turns out that the proper
definition, precisely in analogy to (\ref{ga}), is given by
\be \label{ga2}
(\eta_F T_{\gamma,\vec{j},\vec{c}})(T_{\gamma',\vec{j}',\vec{c}'}):=
\int_{\Rl^n} \frac{d^nt}{(2\pi)^n}
<T_{\gamma,\vec{j},\vec{c}},U_{\gamma\cup\gamma'}(t_1,..,t_n)
T_{\gamma',\vec{j}',\vec{c}'}>
\ee
where $n=n(\gamma\cup\gamma')$. Namely, using the definition of $\mu_0$
which assigns to each holonomically independent loop one independent 
integration variable with respect to the Haar measure on $SU(2)$ we 
explicitly compute that (\ref{ga2}) equals 
\be \label{ga3}
\int d\mu_H(g_1)..d\mu_H(g_m)[\overline{T_{\gamma,\vec{j},\vec{c}}}
T_{\gamma',\vec{j}',\vec{c}'}](g_1,..,g_m)
=:\int_{\cal M} d\nu(A_0)
(\overline{T_{\gamma,\vec{j},\vec{c}}}
T_{\gamma',\vec{j}',\vec{c}'})(A_0)
\ee
where the square bracket on the left hand side means that the function 
is to be evaluated on the trivial holonomy for the contractable loops 
which thus leaves only an integration over holonomies $g_1,..,g_m$ along 
loops that generate the homotopy group of $\Sigma$. The right hand side 
defines the measure $d\nu$ on $\cal M$ and agrees with the measure
given by Witten \cite{Witten}. It is easy to see that (\ref{ga2})
coincides with (\ref{rigger}).   Note that, even though we must make
a choice of generators of $\pi_1(\gamma \cup \gamma')$ to even write
down the integral (\ref{ga2}), the resulting definition of 
$\eta_F$ is independent of this choice.
In addition, note that we have seen no sign of the superselection rules 
that arose in \cite{jmp}.  We shall return to this issue in the
next section.

\section{A Solution in two Stages}
\label{Diff}

Recall that one of the main objectives of the present paper is to 
solve the theory using the space of diffeomorphism invariant
states (from \cite{jmp}) as an intermediate step.  That is, 
we use a rigging map $\eta_{Diff}$ from \cite{jmp} to define a
Hilbert space $H_{Diff}$ of diffeomorphism invariant states
and then solve the Hamiltonian constraint\footnote{In the 3+1
case there are some additional difficulties
with such an approach due to the fact 
that the corresponding Hamiltonian constraint does not commute with 
diffeomorphisms. 
In the present model 
this problem does not occur because the set of $F=0$ constraints is invariant 
under diffeomorphisms.} using a second topological vector subspace 
$\Phi_{Diff}$ of 
$H_{Diff}$ and a rigging map $\eta_{Ham}: \Phi_{Diff}
\rightarrow \Phi'_{Diff}$.  That is, roughly speaking, we wish to
write
\begin{equation}
\label{comp}
\eta_F = \eta_{Ham} \circ \eta_{Diff}.
\end{equation}
In contrast, in section 3 we solved all of the constraints
in one step. As outlined in the introduction, the diffeomorphism 
and Hamiltonian constraint are included in the $F=0$ constraint. What we 
would like to see now is how the $F=0$ constraint can be split into two
parts, the diffeomorphism part and a remainder.  This remainder will, 
in some sense, define 
our `Hamiltonian constraint'.

There are, however, two immediate problems with (\ref{comp}).
The fist is that each rigging map is anti-linear, so that the left hand
side is anti-linear while the right is linear.  The other is
that the left hand side is a map from $\Phi_{Cyl}$ to $\Phi_{Cyl}'$, while
the right is a map from $\Phi_{Cyl}$ to $\Phi_{Diff}'$ (through
$\Phi_{Diff}$).  Clearly then, we will need a natural anti-linear map
$\sigma: \Phi'_{Diff} \rightarrow \Phi_{Cyl}'$.  This map will be an
extension of an adjoint map, and will be discussed below
in the course of our argument.

We do this as follows. Recall that each diffeomorphism invariant
distribution in the space $\Phi_{Diff}$(constructed in \cite{jmp})
is a linear combination of spin-network states 
associated with a finite number of graphs. To be more 
precise, collect the triple $\gamma,\vec{j},\vec{c}$ into a single index
$c$ and let $T_{[c]}(A)$ be the distribution defined by
\be \label{DISNS}
T_{[c]} (A):=\sum_{c'\in [c]} T_{c'} (A)
\ee
where $[c]$ is the set of labels of the spin-network states
that one obtains by acting on $T_c$ with
all possible analytic diffeomorphisms. Our objective is now 
to write a solution $\delta(F) T_c$ to the $F=0$ constraint in terms of
$\eta_{Diff}$ and a remaining operation $\eta_{Ham}$ to be obtained.
To that end we write (\ref{modspace}) explicitly as 
\be \label{simple1}
\delta(F)=\sum_c \overline{T_c(A)} \int_{\cal M} d\nu(A_0) T_c(A_0)
=:\sum_c \overline{T_c(A)} k_c
\ee
where the sum is over {\em all} labels $c$.
What helps us now is that since $T_c(A_0)$ is diffeomorphism invariant
for $A_0\in {\cal M}$, it follows that the integrals $k_c$ do not depend 
on $c$ but only on the diffeomorphism equivalence class $[c]$.  We also
note that $T_c(A)$ is real, so that we may drop the overline.

We will therefore relabel $k_c$ as $k_{[c]}$ 
and so write (\ref{simple1}) in the form 
\be \label{simple2}
\delta(F)=\sum_{[c]} k_{[c]}{T_{[c]}(A)} 
\ee
which is already a sum of diffeomorphism invariant distributions only.

If we introduce the notation $T_{[c]}$ for the linear functional on
$\Phi_{Cyl}$ given by $T_{[c]}(\phi) = \int_{\agb} T_{[c]}(A) \phi(A)$, 
then we may write
\begin{equation}
\eta_F(1) = \sum_{[c]} k_{[c]} T_{[c]}.
\end{equation}

In order to connect with \cite{jmp}, recall that, due to the superselection
rules, $\eta_{Diff}$ was not uniquely defined in \cite{jmp}.  In fact, 
the possible rigging maps were labeled by uncountably many real parameters.
However, all of these maps were of a similar form.  Let us simply choose
one of these maps and refer to it as $\eta_{Diff}$.  We will see that
nothing will depend on which map was chosen.  Note that $\eta_{Diff}$
then has the form
\begin{equation}
\eta_{Diff} T_c = \alpha_{[c]} T_{[c]}
\end{equation}
for some $\alpha_{[c]} \in {\R^+}$.

Let $c_0$ be a particular label. We wish to place our
rigging map in the form
\be \label{combine}
\eta_F T_{c_0}=:(\sigma \circ \eta_{Ham} \circ \eta_{Diff})T_{c_0}.
=(\sigma \circ \eta_{Ham}) T_{[c_0]} \alpha_{[c_0]}
\ee
The map $\eta_{Ham}$ will act on
$\Phi_{Diff}$, the group averaged cylindrical functions on $\agb$. 
Notice that the space $\Phi_{Diff}$ is a space of distributions on
$\Phi_{Cyl}$ but a space of test functions for the space $\Phi_{Diff}'$,
the dual of $\Phi_{Diff}$. 

We now address the map $\sigma$.  It is to be an anti-linear map
from $\Phi_{Diff}'$ to $\Phi_{Cyl}'$.  We will construct this
map by (anti)linearly extending the adjoint map on ${\cal H}_{Diff}$.
Recall that, ${\cal H}_{Diff}$ is defined through the following 
inner product on $\Phi_{Diff}$:

\be \label{InvIp}
<\eta_{Diff}\phi,\eta_{Diff}\phi'>_{Diff}:=[\eta_{Diff}\phi](\phi')\;
{\rm for \ all} \;\phi,\phi'\in\Phi_{Cyl}.
\ee
Thus, (\ref{InvIp}) defines an anti-linear (adjoint) map
$\dagger: \Phi_{Diff} \rightarrow \Phi_{Diff}'$.  On the image
$\dagger \Phi_{Diff} \subset \Phi_{Diff}'$, this map is invertible and
the inverse $\dagger^{-1}$ is also anti-linear.  We note that $\dagger
\Phi_{Diff}$ in fact provides a basis for $\Phi_{Diff}'$ and that
$\Phi_{Diff} \subset \Phi_{Cyl}'$.  Using anti-linearity then, we
may attempt to extend $\dagger^{-1}$ to a map from all of 
$\Phi_{Diff}'$ into $\Phi_{Cyl}'$.  The result is in fact well defined
and gives the desired map $\sigma: \Phi_{Diff}' \rightarrow \Phi_{Cyl}'$.

Let us now define $\eta_{Ham}$ to be of the form
\be \label{ansatz}
\eta_{Ham}\cdot T_{[c]}=\sum_{[c']} a([c],[c']) T_{[c']}^\dagger,
\ee
where $\dagger: \Phi_{Diff} \rightarrow \Phi_{Diff}'$ is the map given
above. Then
\begin{equation}
(\sigma \circ \eta_{Ham}) T_{[c]} = \sum_{[c']} a^*([c],[c']) T_{[c']}.
\end{equation}
The coefficients $a$ will be chosen so that 
(\ref{combine}) is satisfied.
The equality 
$\alpha_{[c_0]} (\sigma \circ
\eta_{Ham})T_{[c_0]}=\delta(F) T_{c_0}$ is to be understood in the sense 
of distributions on $\Phi_{Cyl}$ and so can be checked by
evaluating both sides on all possible $T_c$. In order to do that we need the
Clebsh-Gordon formula
\be \label{CG}
T_{c_0} T_c=\sum_{c'} b(c_0,c;c') T_{c'}
\ee
which is a finite sum thanks to the piecewise analyticity of the graphs 
involved. Notice that the coefficients $b(c_0,c;c'(c_0,c))$ are invariant 
under simultaneous diffeomorphic mappings of $c_0$ and $c$.\\

Finally, using 
${T_{[c]}}(T_{c'})=\chi_{[c]}(c')$ (where $\chi_{[c]}(c')$ is the
characteristic function given by $1$ for $c' \in [c]$ and
 $0$ otherwise) together with (\ref{CG}), we find
\ba \label{evaluate}
[\eta_FT_{c_0}](T_c)&=&[\eta_F(1)](T_{c_0} T_c)\nonumber\\
&=& \sum_{c'} b(c_0,c;c') [\eta_F(1)](T_{c'})=\sum_{c'}b(c_0,c;c') k_{[c']}
\nonumber\\
\ea
The first equality in (\ref{evaluate}) uses the 
fact that $T_c$ is real valued. Notice that despite the appearance 
of the Clebsh-Gordon coefficients
$b(c_0,c;c')$ (which seem to depend on $c_0,c$),
the corresponding sum actually 
depends only on the equivalence classes $[c_0],[c]$ since
we have $[\varphi\cdot T_{c_0}][\varphi'\cdot 
T_c]=T_{c_0}T_c$ on the space of flat connections,
for arbitrary
$\varphi,\varphi'\in\mbox{Diff}(\Sigma)$.

Thus, if we define $a([c_0],[c]) = \alpha^{-1}_{[c_0]}
\sum_{c'}b(c_0,c;c') k_{[c']}$, then, since the Clebsh-Gordon
coefficients are real
we have
\begin{equation}
[(\sigma \circ \eta_{Ham} \circ \eta_{Diff}) (T_{[c_0]})](T_c) = 
\sum_{[c']} b([c_0],[c];c') k_{[c']} =
[\eta_{Ham} T_{[c_0]}](T_c).
\end{equation}
That is, we have constructed a map $\eta_{Ham}: \Phi_{Diff}
\rightarrow \Phi_{Diff}' \subset \Phi_{Cyl}'$ such that $\eta_F
= \sigma \circ \eta_{Ham} \circ \eta_{Diff}$.  Note that the composition
$\eta_{Ham} \circ \eta_{Diff}$ is independent of 
$\alpha_{[c]}$, and thus independent of the particular choice of the map
$\eta_{Diff}$.

Let us now examine the status of the `superselection rules'
described in \cite{jmp}, associated with the averaging over diffeomorphisms.
According to these rules, the states $T_{[c]}, T_{[c']} \in
H_{Diff}$ were superselected whenever $c$ and $c'$ were associated
with graphs $\gamma,\gamma'$ in distinct diffeomorphism classes.
Note, however, that in this case we may still have
$[\eta_{Ham}(T_{[c]})](T_{[c']}) \neq 0$ or, equivalently,
\begin{equation}
\langle T_{[[c]]},T_{[[c']]}\rangle_{phys}  \neq 0.
\end{equation}
So that the corresponding states are not superselected.  In fact, whenever
$c$ and $c'$ are associated with homotopically equivalent
triples $c= (\gamma,\vec{j}, \vec{c})$ and
$c'= (\gamma',\vec{j}{}', \vec{c}{}')$, the
states $T_{[[c]]}$ and $T_{[[c']]}$ are proportional.  Furthermore, 
the operator $\hat{T}_{\alpha,phys}: \hat{T}_{\alpha,phys} \eta_F f
= \eta_F [T_\alpha f]$ is well defined and mixes even homotopically 
distinct graphs.  As a result, no sign of the superselection rules remains
in the physical Hilbert space.

\end{section}
\begin{section} {Conclusions}

\label{concl}

We have seen in section 2 that the `superselection rules'
among the diffeomorphism invariant states in no way carry over to
the physical space.  This result was not unexpected, as we
took the `correct' description of 2+1 Euclidean gravity to be that
given by Witten \cite{Witten} in which no superselected sectors arise.
However, the appearance of such spurious superselection rules in an
intermediate stage was in no way an obstacle to the solution of the theory
using loop representation and refined algebraic techniques, or even
to solving first the diffeomorphism constraint and then implementing
the remaining constraints.  We recall that any of the 
possible maps $\eta_{Diff}$ can be used and that they all lead
to the same physical Hilbert space in the end\footnote{In our
presentation this was true by construction.  However, the
map $\eta_F$ is unique at least up to the choice of measure $d\nu$ while
any choice of $\eta_{Diff}$ is compatible with any choice of $d\nu$.
Therefore, given any $\eta_{Diff}$, the same array of physical
Hilbert spaces may be constructed through maps of the form
$\eta_{Ham} \circ \eta_{Diff}$.}. 
This can be taken as an encouraging sign
for a similar approach in the 3+1 case.  On the other hand, we have used the 
Witten constraints and the fact that they are well defined on $H_{aux}$
to achieve our goals.  Such techniques are not available in the 3+1
case; it remains to be seen if this difference is crucial.

Let us now address the question of whether the diffeomorphism superselection
rules will be spurious in 3+1 gravity.  We first note that, 
as described in \cite{jmp,ban},
there are many examples for which superselection laws arising from
RAQ {\it are} of physical relevance, as they have analogues even
at the classical
level.  What accounts for the difference between these systems?
Several answers may be given.  For example, in 
\cite{ban} it was found that spurious superselection rules can arise
through a poor choice of the subspace $\Phi$.  In general though, 
it appears that such spurious superselection rules are associated
with singular structures in either the system or in our description of it.

To illustrate this point, recall the RAQ deals directly with only
the {\it strong} observables of the system.  Now, at least classically, 
this is no problem for any sufficiently smooth system.  
Let $\Gamma$ be the phase space of a classical system with $C$ the
corresponding constraint surface and $G$ the group of gauge transformations.
When $C/G$ is a smooth submanifold of $\Gamma/G$, all of the physics
is indeed captured by strong observables.  Any observable (that is, 
any function on $C/G$) may be extended to $\Gamma/G$ and pulled
back to $\Gamma$, where it defines a strong observable.   Thus, the
strong observables capture all of the physics of the system.

In our quantum case, however, there were interesting observables
(the $\hat{T}_{\alpha,phys}$) which were not {\it strongly} diffeomorphism
invariant.  It would be interesting to understand whether this was
due to some sort of singularity in the classical phase space or simply due
to our quantum description.  In any case, something analogous happens
for 3+1 systems.  It is shown in Appendix A that, 
at least in the
representation based on the Ashtekar-Lewandowski Hilbert space, 
there are many quantum operators which are weak observables (with
respect to the diffeomorphism constraint) but which do not become
equivalent to
any strongly diffeomorphism invariant observable when the diffeomorphism
constraints are imposed.  

There are of course several possible interpretations here.  Note that
Appendix A
considers only the diffeomorphism (and gauge) constraints.
It is therefore possible that, once the full
algebra including the Hamiltonian constraints are considered, no
analogue of these operators will remain.  Another possibility
is that these observables are simply spurious results of the quantization
method and have no physical meaning.  A third, however, is that such
observables are important for a proper treatment of the system and that
we must expand our techniques to take them into account.  In any case, 
when we consider that the Hamiltonian constraint of 3+1 gravity
\cite{TT} mixes the `superselected sectors' much as
$\eta_{Ham}$ does in the 2+1 case, it appears likely that the
diffeomorphism superselection laws are spurious in 3+1 gravity as well.
\end{section}

\centerline{\bf Acknowledgments}

Many thanks to the ESI, where much of this work was accomplished, for
their hospitality and partial support.  DM was supported in part
by NSF grant PHY95-07065 and funds provided by Syracuse University.
JM was supported in part by CERN grant CERN/S?FAE/1030/95 and in
part by CENTRA/Algarve.  TT was supported in part by DOE
grant DE-FG02-94ER25228 to Harvard University.
We would like to thank Abhay Ashtekar for suggesting this project and
Abhay Ashtekar, John Baez, and Jerzy Lewandowski for useful discussions.

\appendix

\begin{section} {Strong Quantum Observables in 3+1}

In this appendix we show that the quantization scheme for 
3+1 gravity considered in \cite{jmp} contains operators that are
weakly invariant under diffeomorphisms but which are not weakly
equivalent to any operator which is {\it strongly} invariant under
diffeomorphisms.
Before proceeding, we should recall certain subtleties of the 
Ashtekar-Lewandowski Hilbert space and carefully define what we
mean by weak observables.  Recall that the diffeomorphism
constraints themselves are not actually defined as operators
on this space \cite{jmp}.  Instead, it is the finite diffeomorphisms (which may
be interpreted as exponentiated versions of the constraints) which
are defined on $H_{aux}$.  These operators are, however, 
sufficient to define a space $\Phi_{Diff}$ of diffeomorphism invariant
states which are naturally thought of as the quantum analogue of the
classical space of solutions to the diffeomorphism constraints.
By weak equivalence of two operators
$B$ and $C$, we therefore mean that $B$ and $C$ coincide when
acting on $\Phi_{Diff}$.  Furthermore, a weak observable is
naturally defined to be one which maps $\Phi_{Diff}$ into itself.

The construction of the observables is quite straightforward.
Recall that $\Phi_{Diff}$ is in fact a space of `dual states,' 
specifically, of linear functionals on the space $\Phi_{Cyl}$
of cylindrical functions.  Thus, any operator $A$ whose
adjoint $A^{\dagger}$ acts on and preserves the space $\Phi_{Cyl}$ of
cylindrical states has a natural `dual action' on $\Phi_{Diff}$
given by
\begin{equation}
[ A \psi_{Diff}](\phi) = \psi_{Diff} (A^\dagger \phi ).
\end{equation}
Now, simply
choose any two nontrivial spin network states $T_1, T_2$ and consider the
sets $S_{T_1}, S_{T_2}$ of all states that can be obtained from $T_1, T_2$ 
respectively by diffeomorphisms.  Since $T_1$ and $T_2$ are each associated
with analytic graphs, the cardinality of both sets $S_{T_1}$, and $S_{T_2}$
is the same, namely that of the power set ${\cal P}({\bf R})$
of all real numbers.  As a result, there is a bijection $\alpha$ between
$S_{T_1}$ and $S_{T_2}$ and our observable $A$ may be defined by
\begin{eqnarray}  
A T = \alpha(T) \ \ & {\rm for} \ T \in S_{T_1} \cr
{\rm while} \ \ A \psi = 0 \ \  \ & {\rm if}
\  \langle \psi| T \rangle = 0 \ {\rm for \ all}
\ T \in S_{T_1}.
\end{eqnarray}
Note that $A$ is a bounded operator whose range
lies in $V_{T_2}$, the space of states spanned by spin network states
in $S_{T_2}$.  The adjoint $A^\dagger$
is of a similar form but is defined by the map $\alpha^{-1}$.  

Both $A$ and $A^\dagger$ are in fact weak observables.
To see this, we simply compute the action of a diffeomorphism on
$A \psi_{Diff}$ for a diffeomorphism invariant state $\psi_{Diff}$.
Since $\Phi_{Diff}$ is a space of linear functionals on $\Phi_{Cyl}$, 
$A \psi_{Diff}$ is entirely determined by its action on spin network
states, which form a basis for $\Phi_{Cyl}$.  For any diffeomorphism $D$ and
any spin network $T$, 
$[D A \psi_{Diff}] (T) = \psi_{Diff} ( A^\dagger D^{-1} T)$.  If
$T$ is orthogonal to the space $V_{T_2}$ (spanned by spin networks
in $S_{T_2}$) then this vanishes.  Otherwise, we may take $T$ to be
$D'T_2$ for some diffeomorphism $D'$.  In either case we have
\begin{equation}
[D A \psi_{Diff}] (T) = \psi_{Diff} ( D''  A^\dagger T) = [A \psi_{Diff}]
(T)
\end{equation}
for some diffeomorphism $D''$.  Thus, $A$ preserves $\Phi_{Diff}$
and is a weak observable with respect to diffeomorphisms.  However, 
since we are free to choose $T_1$ and $T_2$ from different
superselected sectors (as defined by \cite{jmp}) for the algebra of
strongly diffeomorphism invariant operators, it is clear that
the action of $A$ on $\Phi_{Diff}$ does not preserve the superselection
sectors.  As a result, $A$ cannot be weakly equivalent to
any strongly diffeomorphism invariant operator.

\end{section}

\end{document}